\documentclass[12pt,preprint]{aastex}

\begin{document}

\shortauthors{Stanek et al.}

\shorttitle{Two Anomalous GRB Afterglows at $z\sim 4$}

\title{``Anomalous'' Optical GRB Afterglows are Common: Two $z\sim 4$
Bursts, GRB 060206 and 060210\altaffilmark{1}}

\author{K.~Z.~Stanek\altaffilmark{2},
X.~Dai\altaffilmark{2},
J.~L.~Prieto\altaffilmark{2},
D.~An\altaffilmark{2},
P.~M.~Garnavich\altaffilmark{3},
M.~L.~Calkins\altaffilmark{4},
J.~Serven\altaffilmark{5},
G.~Worthey\altaffilmark{5},
H.~Hao\altaffilmark{6},
A.~Dobrzycki\altaffilmark{7},
C.~Howk\altaffilmark{3},
T.~Matheson\altaffilmark{8}
}

\altaffiltext{1}{Based on data from the MDM 2.4-m and 1.3-m telescopes}
\altaffiltext{2}{Dept. of Astronomy, The Ohio State University, Columbus, OH 43210}
\altaffiltext{3}{Dept. of Physics, University of Notre Dame, Notre Dame, IN 46556}
\altaffiltext{4}{F.~L.~Whipple Observatory, Amado, AZ 85645}
\altaffiltext{5}{Dept. of Physics, Washington State University, Pullman, WA 99164}
\altaffiltext{6}{Harvard-Smithsonian Center for Astrophysics,Cambridge, MA 02138}
\altaffiltext{7}{European Southern Observatory, Garching bei M\"unchen, Germany}
\altaffiltext{8}{NOAO Gemini Science Center, Tucson, AZ 85719} 

\email{kstanek@astronomy.ohio-state.edu}

\begin{abstract}

We report on two recent $z\sim 4$ gamma-ray bursts (GRBs), GRB\,060206
and GRB\,060210, for which we have obtained well-sampled optical light
curves. Our data, combined with early optical data reported in the
literature, shows unusual behavior for both afterglows. In $R$-band
GRB\,060206 ($z=4.045$) experienced a slow early decay, followed by a
rapid increase in brightness by factor $\sim 2.5$ about 1 hour after
the burst.  Its afterglow then faded in a broken power-law fashion,
with a smooth break at $t_{b}=0.6\;$days, but with additional, less
dramatic ($\sim 10\%$) ``bumps and wiggles'', well detected in the
densely sampled light curve. The $R$-band afterglow of GRB\,060210
($z=3.91$) is also unusual: the light curve was more or less flat
between 60 and $300\,$sec after the burst, followed by $\sim 70\%$
increase at $\sim 600\,$sec after the burst, after which the light
curve declined as a $\sim t^{-1.3}$ power-law.  Despite reports to the
contrary, we find that for GRB\,060206 X-rays follow the optical
decay, but with significant variations on short timescales. However,
the X-ray afterglow is contaminated by a nearby, variable source,
which especially at late times obscures the behavior of the X-ray
afterglow. The early X-ray light curve of GRB\,060210 exhibited two
sharp flares, without corresponding peaks in the optical light. We
find that the late X-ray light curve is well described by a broken
power-law model, with a smooth break at $\sim 8\;$hours after the
burst. The early decay index of the X-ray light curve is not
consistent with the $\sim t^{-1.3}$ power-law seen in optical.

We argue that ``anomalous'' optical afterglows are likely to be the
norm, and that the rapid variations often seen in {\em Swift}-XRT data
would also be seen in the optical light curves, given good enough
sampling.  As a result, some of the often employed procedures, such as
deriving the jet opening angle using a broken power-law fit to the
optical light curves, in many cases might have a poor statistical
significance.  Finally, we argue that the rapid rise at $\sim
3000\;$sec in the optical for GRB\,060206 and the optical bump at
$\sim 700\;$sec in GRB\,060210 might be due to the turn-on of the
external shock.  If indeed the case, the existence and timing of such
features could provide valuable additional information about the
energy of the GRB jet and the density of the circumburst medium it is
plowing into.

\end{abstract}


\section{Introduction}

Gamma-ray bursts (GRBs) continue to surprise us. With the {\em Swift}
satellite \citep{gehrels04} providing quick and accurate localizations
for many bursts (for example, 20 GRBs were localized between January
1st - February 15th, 2006), it is now possible to find and study in
even more detail bursts which are ``unusual''. Those are more than
just mere curiosity---they test and verify our understanding of
physics of these extreme events.

The physics of GRB afterglow emission appears on solid footing. The
GRB sweeps up ambient gas and the resulting shock emits synchrotron
radiation from X-rays to radio wavelengths
\citep[e.g.][]{Piran05}. The light curve decays as a power-law with a
break when the opening angle of the beamed emission exceeds the
opening angle of the jet (e.g., Stanek et al.~1999; Rhoads 1999). But
the fast localizations of the {\em Swift} satellite means that optical
and X-ray observations can begin before the $\gamma$-rays have
completely faded. The transition between the end of the ``prompt''
emission and the rise of the shocked ambient gas emission is uncertain
territory.  Internal shocks may be the source of prompt emission or
continued activity from the central engine may inject more energy into
the external shock. Deciphering the effects of these processes
requires high-quality observations obtained rapidly after the burst.
What also helps is time dilation. High-redshift bursts have the
advantage of being observed in slow motion.

Two such high-redshift bursts have been recently detected by the {\em
Swift} satellite. GRB\,060206 triggered {\em Swift}-BAT on Feb. 6,
04:46:53 UT \citep{morris2006a}. A likely afterglow has been
identified by \citet{fynbo06a}, who also determined that the afterglow
was at high redshift of $z=4.045$ \citep{fynbo06b}. A presence of a
bright afterglow has been reported by several groups, most notably by
RAPTOR \citep{wozniak06a}, who indicated that the afterglow has
increased in brightness by $\sim 1\;$mag about 1 hour after the burst.
\citet{wozniak06b} has released detailed description of the RAPTOR
data for this event, which we will discuss later in this paper.

GRB\,060210 triggered {\em Swift}-BAT on Feb. 10, 04:58:50 UT
\citep{Beardmore06a}. \citet{fox06} have quickly identified a possible
afterglow, and \citet{Cucchiara06} measured a high redshift of
$z=3.91$. KAIT robotic telescope has obtained extensive optical
observations starting as soon as 62 sec after the burst \citep{Li06a},
indicating that the afterglow brightened by 0.4 mag by 9 minutes after
the burst \citep{Li06b}. \citet{Beardmore06b} reported on the {\em
Swift}-XRT data, which showed two strong flares 200 sec and 385 sec
after the trigger, followed by a power-law decay.

In this paper we report on our optical follow-up of these two
bursts. Our data, described in Section 2, combined with the data
reported in the literature clearly shows two very unusual optical
afterglows, whose evolution we describe in Section 3. In section 4 we
analyze {\em Swift}-XRT data for these two events. We briefly
summarize our results and discuss their implications in Section 5.

\section{The Optical and X-Ray Data}

The majority of our data were obtained with the MDM-2.4m telescope,
with additional data obtained using the MDM-1.3m
telescope. For the bright afterglow of GRB\,060206, we have
obtained 83 high signal-to-noise $R$-band images between 1.7 and 8.7
hours after the burst, followed by additional 16 $R$-band images
during the next two nights. For the much fainter afterglow of
GRB\,060210, we have obtained 12 $R$-band images between 0.48 and 2.0
hours after the burst, until the object has set.

All the light curves were extracted using ISIS2 image subtraction
package \citep{alard00}. To obtain absolute calibration, for
GRB\,060206 we used nine stars in the field with SDSS photometry
\citep{cool06,adelman06} to transform to the standard system fitting a
zero-point difference and a color-term. We used the transformation
equations of \citet{lupton05} to transform SDSS griz magnitudes to
BVRI. For future references and cross-calibrations, this
transformation gives $R=15.89 \pm 0.02\;$mag and $I=15.47 \pm
0.02\;$mag for the star SDSS J133130.4+350416.1. We assumed
$R-I\simeq0.6\;$mag for the afterglow \citep{greco06}. The absolute
photometric calibration is thought to be better than $\sim$5\%. We
found that in order to match \citet{wozniak06b} data to our data, we
had to subtract $0.22\;$mag from their unfiltered $R$-band magnitudes
(see Fig.2). This offset does not affect any of our conclusions. For
GRB\,060210 we used the calibration of the field obtained with the
KAIT telescope (Li et al. 2006, in preparation).  A USNO-B1
\citep{monet98} star 1170-0048923 at ${\rm RA}=\;$03:50:53.00, ${\rm
DEC}=+$27:01:30.54 (J2000), which is close to the GRB, is measured at
$R=15.64\;$mag using KAIT calibration, $0.29\;$mag fainter than the
USNO magnitude. This is not unusual, since the magnitudes of stars in
the USNO-B1 catalog were obtained with photographic plates. Tables~1
and 2 present our $R$-band photometry for the two bursts.
 
We have also analyzed the X-ray data obtained by the {\em Swift}-XRT
instrument. We started with the XRT level 2 event files for both the
windowed timing mode and the photon counting mode observations. There
was a hot CCD column in the windowed timing observation for
GRB\,060206.  We filtered the events on these hot CCD pixels. We used
the
\verb+xselect+\footnote{http://heasarc.gsfc.nasa.gov/docs/software/lheasoft/ftools/xselect/xselect.html}
software package to extract the X-ray light curves and spectra.  The
background subtracted light curves were adaptively binned according to
the signal-to-noise ratios. We fit the spectra with \verb+XSPEC+
(Arnaud 1996) to convert XRT count rates to fluxes. We use the
\verb+rmf+ files from standard XRT calibration distribution, and
generated the \verb+arf+ files with the {\em Swift}-XRT software tool
\verb+xrtmkarf+\footnote{http://swift.gsfc.nasa.gov/docs/swift/analysis/xrt\_swguide\_v1\_2.pdf}
.

\section{Evolution of Optical Afterglows}

\begin{figure}
\plotone{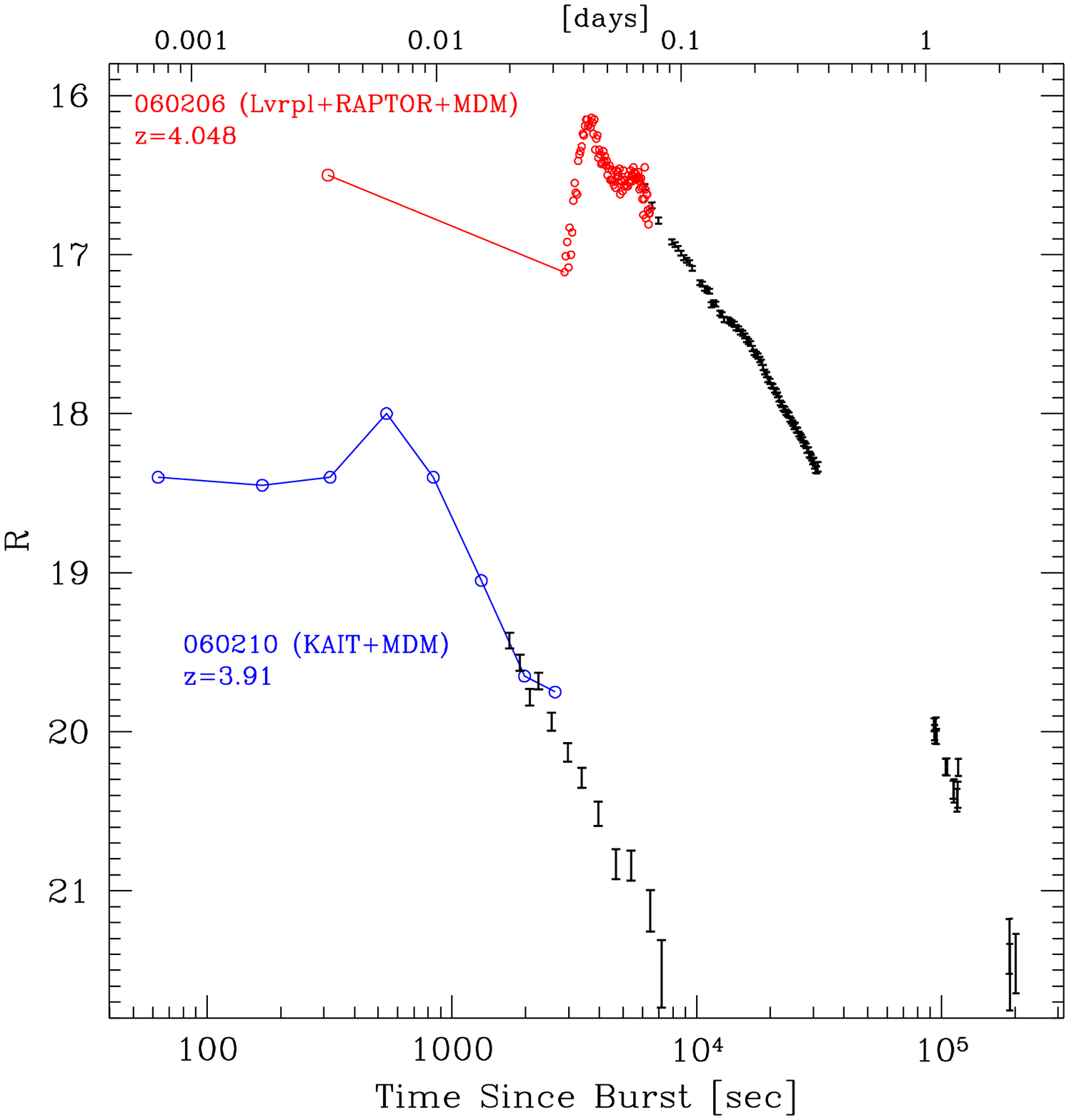}
\caption{$R$-band light curves of two $z\sim 4$ afterglows,
GRB\,060206 and 060210. Our data are shown with the error bars, while
the data adopted from the literature (GRB 060206: Guidorzi et
al. 2006, Wozniak et al. 2006b; GRB 060210: Li et al. 2006b) are shown
with the open points, connected together for clarity. }
\label{fig_R}
\end{figure}

Figure~\ref{fig_R} presents the $R$-band light curves for GRB\,060206
and GRB\,060210 (see Tables 1,2). In addition to our MDM 2.4-m data,
we have added some data from the literature to extend the time
coverage. For GRB\,060206, we add one early data point from
\citet{guidorzi06} and 101 densely sampled RAPTOR data points from
\citet{wozniak06b}. For GRB\,060210, we add eight very early data
points from the KAIT telescope reported by \citet{Li06a, Li06b}. All
the literature data were brought to the common zero-point with our
data, using overlaps between the data sets.

\begin{figure}
\plotone{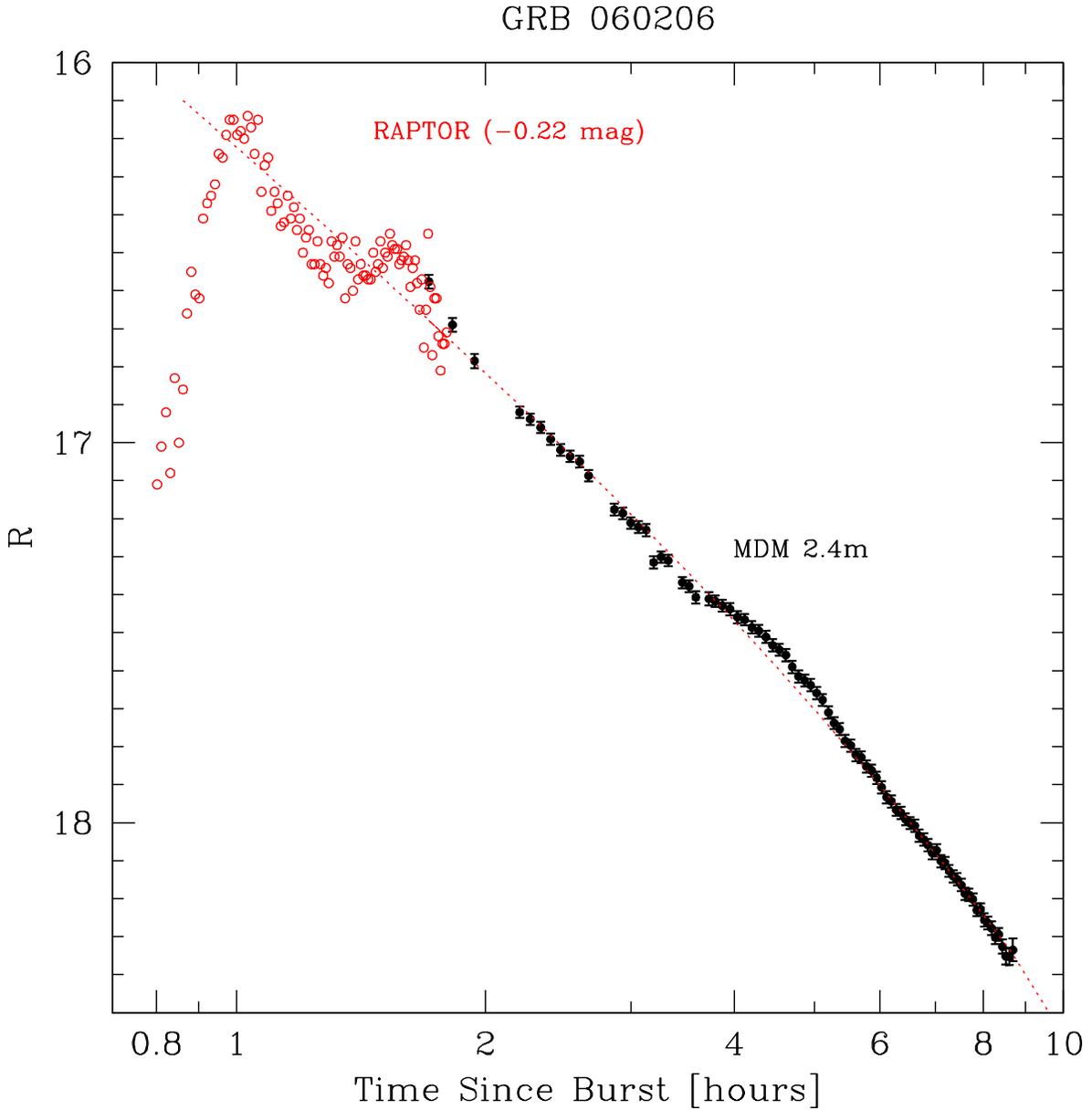}
\caption{$R$-band light curve of the optical afterglow of GRB\,060206
during $0.8-8.7\;$hours after the burst.  RAPTOR data
\citep{wozniak06b} are shown with open points, shifted by $-0.22\;$mag
to match to our data, which are shown as filled points with
errorbars. A broken power-law fit to all our $R$-band data is shown
with the dotted line.}
\label{fig_bump}
\end{figure}

Figure~\ref{fig_R} shows most unusual behavior for both afterglows. As
described in detail by \citet{wozniak06b}, GRB\,060206 experienced a
slow early decay, followed by a rapid increase in brightness by $\sim
1\;$mag about 1 hour after the burst.  Its afterglow then faded in a
typical broken power-law fashion, with a smooth break at
$t_{b}=0.6\;$days, but with additional, less dramatic ($\sim 10\%$)
``bumps and wiggles'', well detected in the densely sampled light
curves.  These deviations from smooth decay can be seen better in
Figure~\ref{fig_bump}, where we show the RAPTOR data and our first
night MDM data.

To characterize the long timescales behavior of its afterglow, we have
fit all our GRB\,060206 $R$-band data with the broken power-law model
of \citet{Beuermann99} (see also \citealt{Stanek01}):
\begin{equation}
F_{R}(t) =
\frac{2F_{R,0}}{\left[\left(\frac{t}{t_b}\right)^{\alpha_1 s}
+\left(\frac{t}{t_b}\right)^{\alpha_2 s}\right]^{1/s}},
\label{eq_brokenpower}
\end{equation}
where $t_b$ is the time of the break, $F_{R,0}$ is the $R$-band flux
at $t_b$ and $s$ is a parameter which determines the sharpness of the
break, where a larger $s$ gives a sharper break.  This formula
smoothly connects the early time $t^{-\alpha_1}$ decay ($t\ll t_b$)
with the later time $t^{-\alpha_2}$ decay ($t\gg t_b$).  The fit
results in the following values for the parameters: $\alpha_1=0.7,
\alpha_2=2.0, t_b=0.6$~days. Given that the data show clear variations
from the smooth behavior, these are only approximate values, and
should be treated with caution (see the Discussion below).  The
overall fit is reasonable and it supports a break in the light curve,
traditionally interpreted as a jet break \citep[e.g.][]{Stanek99}.
This is broadly consistent with the behavior in other optical bands
reported by \citet{lacluyze06} and \citet{reichart06}.

The $R$-band afterglow of GRB\,060210 ($z=3.91$) closely resembles
recently reported behavior of GRB\,0508101 \citep{rykoff06}, who
called that event an ``Anomalous Early Afterglow''. The light curve of
GRB\,060210, as reported by \citet{Li06b}, was more or less flat
between 60 and $300\,$sec after the burst, followed by $\sim 0.4\;$mag
increase by $\sim 600\,$sec after the burst. Using \citet{Li06b} and
our data we show that the light curve then declined as a $\sim
t^{-1.3}$ power-law. If we bring GRB\,060210 to a fiducial redshift of
$z=1$, which is close to a likely redshift of GRB\,050801
\citep{rykoff06}, then these two burst are even closer in their
evolution, as can be seen comparing to their Figure~1.

\section{Comparison of Optical to X-Ray Data}

Given the unusual behavior observed in the optical wavelengths for
these two bursts, it is useful to investigate their X-ray light curves
as well. Indeed, X-ray afterglows observed by {\em Swift}-XRT have
been shown to have features \citep{nousek06} not expected in the
standard afterglow models, including giant flares such as observed in
GRB 050502B \citep{falcone06}. The origin of the flares is still under
investigation \citep[e.g.][]{zhang06, Perna05}.

\begin{figure}
\plotone{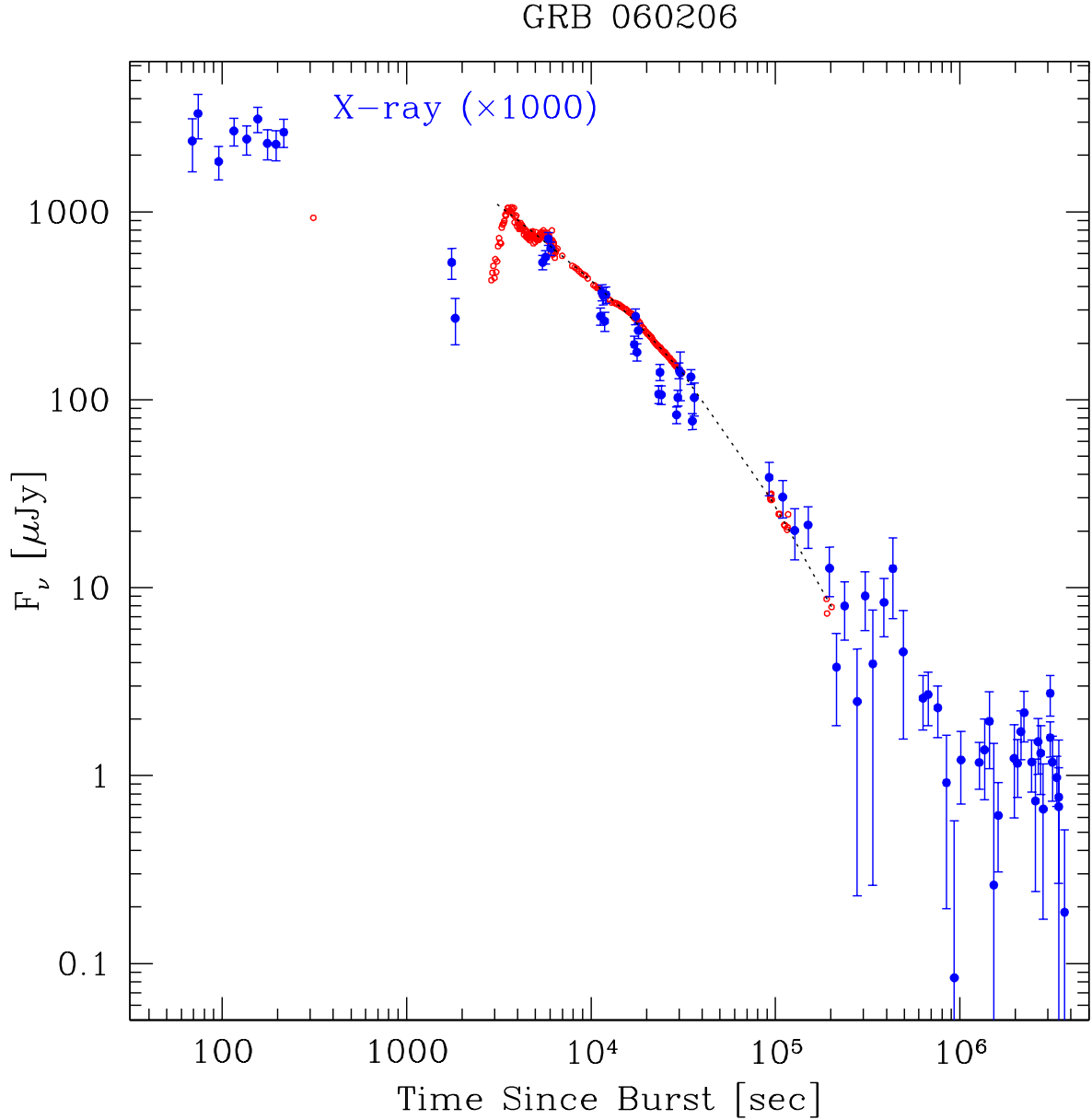}
\caption{Comparison of the $R$-band (open points) to X-ray light curve
(filled points with error bars) for the GRB\,060206. To better compare
to the optical flux, the $F_{\nu}$ for the {\em Swift} X-ray light
curve has been multiplied by a factor of 1000. For the optical data,
with the dotted line we show the broken power-law fit described in the
text.}
\label{fig_xray1}
\end{figure}

In Figure~\ref{fig_xray1} we compare the X-ray to optical light curves
for GRB\,060206. While it was intrinsically a very bright afterglow in
optical, it was a faint X-ray event, with the ratio of
$F_{\nu,R}/F_{\nu,X}\sim 1000$. Despite earlier reports to the
contrary \citep{morris2006b}, the overall behavior between the two
bands is similar, but with clear short-timescale variations, as
reported before by \citet{morris2006b}. However, by analyzing the
later XRT observation, we clearly see a nearby, contaminating X-ray
source, about $15\arcsec$ away. This nearby source is most likely
variable, obscuring the true behavior of the X-ray afterglow,
especially at later times. That also explains the flattening of the
late X-ray light curve, as seen in Figure~\ref{fig_xray1}.

\begin{figure}
\plotone{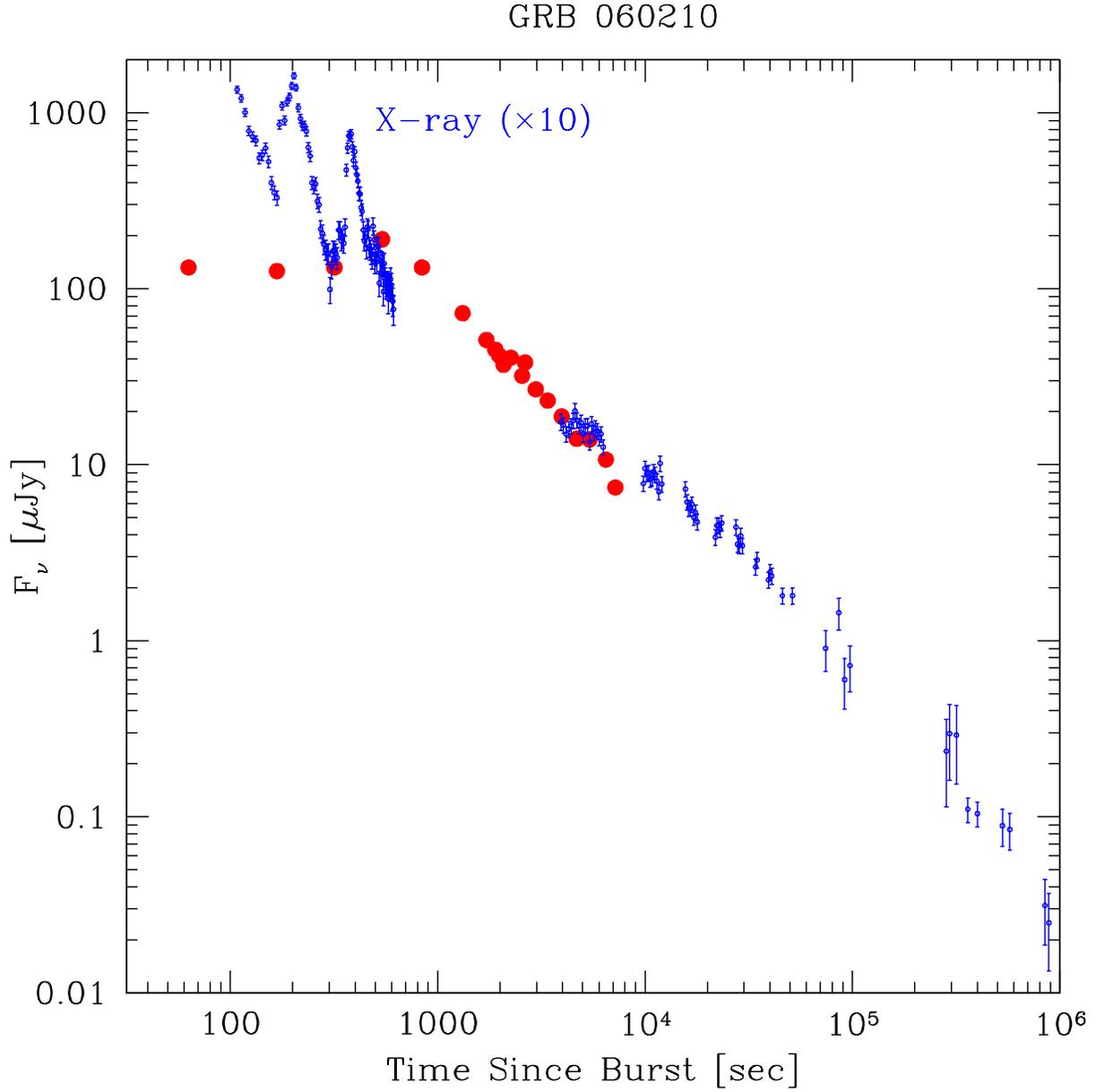}
\caption{Comparison of the $R$-band data (filled points) to X-ray
light curve (error bars) for the GRB\,060210. To better compare to the
optical flux, the $F_{\nu}$ for the {\em Swift} X-ray light curve has been
multiplied by a factor of 10. }
\label{fig_xray2}
\end{figure}

In Figure~\ref{fig_xray2} we compare the X-ray to optical light curves
for GRB\,060210. Here the ratio of $F_{\nu,R}/F_{\nu,X}\sim 10$, much
lower than for GRB\,060206.  As reported by \citet{Beardmore06b},
X-rays show two strong flares 200 sec and 385 sec after the trigger,
followed by a smoother decay. As already reported by Dai \& Stanek
(2006), the analysis of the entire XRT light curve indicates a
presence of a broken power-law.  They fitted both a single power-law
and a broken power-law model to the XRT light curve from $3.0\times
10^{3}\;$sec to $1.0\times 10^{6}\;$sec after the BAT trigger.  For
the single power-law model they found a decay index of $\alpha=1.09$,
but the fit was poor.  For the broken power-law model they found
$\alpha_1 = 0.7, \alpha_2 = 1.4$, $t_b = 7.9\;$hours, with the broken
power-law providing a much better fit.  The power-law decay index for
the X-ray light curve before the jet break is not consistent with the
optical decay index of $\sim 1.3$ discussed above.  It would be most
interesting to add later $R$-band data for this event to see if the
X-rays are indeed different from optical.

We note that the X-ray flares in GRB\,060210 do not have corresponding
optical peaks \citep[as far as we can tell from the data reported
by][]{Li06b}.  This would indicate that the X-ray flares are occurring
in a different region from where the optical light is formed.

\section{Summary and Discussion}

We have presented optical light curves for two recent $z\sim4$
afterglows, GRB\,060206 and GRB\,060210. They both show unusual
behavior, with significant re-brightening by as much as factor of
$\sim 2.5$ about 1 hour ($\sim 10\;$min in the rest frame) after the
burst in case of 060206. GRB\,060210 also shows unusual optical
evolution, similar to ``anomalous afterglow'' of GRB\,050801 recently
described by \citet{rykoff06}. Both bursts show complex behavior in
the X-rays as seen with {\em Swift}-XRT instrument.

Significant ``bumps an wiggles'' have been seen before in a number of
optical afterglows.  One of the first well-observed afterglows,
GRB\,970508 \citep[e.g.][]{galama98} had a light curve rather similar
in shape to GRB\,060206, as it brightened by $> 1\;$mag by $\sim
2\;$days after the burst, after which it decayed is a smooth power-law
fashion.  GRB\,000301C exhibited achromatic, short-timescale bump that
was difficult to reconcile with the standard relativistic shock model,
and which was proposed to be caused by microlensing
\citep{garnavich00}. GRB\,021004 \citep[e.g.][]{Bersier03} was not
only bumpy, but also showed clear color variations in optical bands
\citep{Bersier03, Matheson03a}.  GRB\,030329 was so bumpy
\citep[e.g.][]{Matheson03b} that if not for the spectroscopic
observations \citep[e.g.][]{Stanek03}, we would still argue if it
showed a ``supernova bump'' or not. In fact, a significant deviations
from a smooth decay have been seen in the light curve of GRB\,030329
as late as two months after the burst (so called ``jitter event'':
\citealt{Matheson03b}; Bersier et al. 2006, in preparation).  Bumps
have also been seen in some previous Swift afterglows (e.g.,
GRB\,050525a: Klotz et al. 2005).

Given the above and the two bursts discussed in this paper, it is
becoming clear that the ``unusual'' or ``anomalous'' optical
afterglows might be more of a norm than an exception. And while ``nice
and smooth'' afterglows have been seen as well (e.g. GRB\,990510:
\citealt{Stanek99}; GRB\,020813: \citealt{Laursen03}; GRB\,041006:
\citealt{Stanek05}) in some well-observed cases, the large number of
anomalous optical afterglows can no longer be seen as a small wrinkle
on the standard afterglow model. In fact, unless there is sufficient
data to suggest otherwise, it would be only prudent to assume that any
given afterglow might be anomalous.  As a result, some of the often
employed procedures, such as deriving the jet opening angle using a
broken power-law fit to the optical light curve, in many cases might
have a poor statistical significance and be simply not applicable.

Finally, given the totality of the data presented in this paper, we
believe that the rise at $\sim 3000\;$sec in the optical for
GRB\,060206 and the optical bump at $\sim 700\;$sec in GRB\,060210
might be due to the turn-on of the external shock---that is, the GRB
jet has swept up enough circumstellar/interstellar gas and the
magnetic field has developed enough so the standard afterglow has
turned-on and it starts to dominate the light in both X-rays and
optical. If that is indeed the case, the existence and timing of such
features could provide valuable additional information about the
energy of the GRB jet and the density of the circumburst medium it is
plowing into.  This idea is discussed in more detail by
\citet{rykoff06}, when describing the ``anomalous afterglow'' of
GRB\,050801.

\acknowledgments

We thank the {\em Swift} team, Scott Barthelmy, and the GRB
Coordinates Network (GCN) for the quick turnaround in providing
precise GRB positions to the astronomical community, and we thank all
the observers who provided their data and analysis through the GCN.
We thank Weidong Li for useful comments and for providing us with his
calibration data for the GRB\,060210.  We also thank the anonymous
referee for useful comments. We also thank the participants of the
morning astro-ph/coffee discussion at the Ohio State University for
useful input.

\clearpage

\clearpage

\clearpage

\clearpage

\clearpage

\begin{deluxetable}{lcrr}
\tabletypesize{\footnotesize}
\tablewidth{0pc}
\tablecaption{\sc $R-$Band Light Curve for GRB\,060206}
\tablehead{
\colhead{Time $[days]$} & \colhead{$R$} & \colhead{$\sigma_{R}$} \\}
\startdata
   0.0712 & 16.576 &  0.018\\
   0.0760 & 16.690 &  0.018\\
   0.0809 & 16.785 &  0.019\\
   0.0917 & 16.920 &  0.015\\
   0.0944 & 16.938 &  0.015\\
   0.0972 & 16.960 &  0.015\\
   0.1000 & 16.991 &  0.015\\
   0.1027 & 17.019 &  0.015\\
   0.1055 & 17.036 &  0.015\\
   0.1083 & 17.050 &  0.015\\
   0.1110 & 17.087 &  0.015\\
   0.1193 & 17.176 &  0.016\\
   0.1221 & 17.186 &  0.015\\
   0.1249 & 17.211 &  0.015\\
   0.1276 & 17.222 &  0.016\\
   0.1304 & 17.230 &  0.016\\
   0.1332 & 17.315 &  0.016\\
   0.1359 & 17.301 &  0.015\\
   0.1387 & 17.310 &  0.015\\
   0.1442 & 17.368 &  0.015\\
   0.1470 & 17.378 &  0.016\\
   0.1498 & 17.407 &  0.016\\
   0.1553 & 17.411 &  0.017\\
   0.1581 & 17.417 &  0.015\\
   0.1612 & 17.429 &  0.016\\
   0.1646 & 17.438 &  0.016\\
   0.1681 & 17.459 &  0.016\\
   0.1715 & 17.466 &  0.015\\
   0.1750 & 17.486 &  0.016\\
   0.1785 & 17.495 &  0.016\\
   0.1819 & 17.511 &  0.016\\
   0.1854 & 17.533 &  0.016\\
   0.1889 & 17.545 &  0.016\\
   0.1923 & 17.559 &  0.016\\
   0.1958 & 17.590 &  0.016\\
   0.1992 & 17.615 &  0.016\\
   0.2027 & 17.626 &  0.016\\
   0.2062 & 17.638 &  0.016\\
   0.2096 & 17.659 &  0.016\\
   0.2131 & 17.677 &  0.016\\
   0.2165 & 17.710 &  0.016\\
   0.2200 & 17.738 &  0.016\\
   0.2235 & 17.754 &  0.016\\
   0.2269 & 17.785 &  0.016\\
   0.2304 & 17.797 &  0.016\\
   0.2339 & 17.822 &  0.016\\
   0.2373 & 17.828 &  0.016\\
   0.2408 & 17.852 &  0.016\\
   0.2442 & 17.863 &  0.016\\
   0.2477 & 17.882 &  0.016\\
   0.2512 & 17.907 &  0.016\\
   0.2546 & 17.933 &  0.016\\
   0.2581 & 17.944 &  0.016\\
   0.2615 & 17.966 &  0.016\\
   0.2650 & 17.974 &  0.016\\
   0.2684 & 17.991 &  0.016\\
   0.2719 & 18.001 &  0.016\\
   0.2754 & 18.008 &  0.016\\
   0.2789 & 18.034 &  0.016\\
   0.2823 & 18.044 &  0.016\\
   0.2858 & 18.059 &  0.016\\
   0.2892 & 18.080 &  0.016\\
   0.2927 & 18.073 &  0.016\\
   0.2962 & 18.101 &  0.016\\
   0.2996 & 18.106 &  0.016\\
   0.3031 & 18.127 &  0.016\\
   0.3065 & 18.141 &  0.016\\
   0.3100 & 18.149 &  0.016\\
   0.3135 & 18.164 &  0.016\\
   0.3169 & 18.187 &  0.016\\
   0.3204 & 18.190 &  0.016\\
   0.3238 & 18.202 &  0.016\\
   0.3273 & 18.230 &  0.016\\
   0.3308 & 18.228 &  0.016\\
   0.3342 & 18.256 &  0.017\\
   0.3377 & 18.264 &  0.017\\
   0.3411 & 18.278 &  0.018\\
   0.3446 & 18.302 &  0.018\\
   0.3481 & 18.294 &  0.017\\
   0.3515 & 18.326 &  0.019\\
   0.3550 & 18.351 &  0.022\\
   0.3585 & 18.353 &  0.022\\
   0.3619 & 18.335 &  0.030\\
   1.0802 & 19.957 &  0.041\\
   1.0844 & 20.006 &  0.048\\
   1.0910 & 20.032 &  0.043\\
   1.0993 & 19.952 &  0.042\\
   1.1076 & 20.030 &  0.047\\
   1.2053 & 20.221 &  0.052\\
   1.2137 & 20.222 &  0.054\\
   1.2220 & 20.222 &  0.054\\
   1.2939 & 20.364 &  0.057\\
   1.3023 & 20.373 &  0.073\\
   1.3407 & 20.431 &  0.072\\
   1.3490 & 20.397 &  0.082\\
   1.3556 & 20.224 &  0.055\\
   2.1932 & 21.350 &  0.173\\
   2.2033 & 21.543 &  0.209\\
   2.3331 & 21.458 &  0.187\\
\enddata
\label{060206_lightcurve}
\tablecomments{Table~\ref{060206_lightcurve} is available in its
entirety in the electronic version of the Journal. A portion is shown
here for guidance regarding its form and content.}
\end{deluxetable}

\clearpage

\begin{deluxetable}{lcrr}
\tabletypesize{\footnotesize}
\tablewidth{0pc}
\tablecaption{\sc $R-$Band Light Curve for GRB\,060210}
\tablehead{
\colhead{Time $[days]$} & \colhead{$R$} & \colhead{$\sigma_{R}$} \\}
\startdata
   0.0199 & 19.427 &  0.050 \\
   0.0220 & 19.566 &  0.051 \\
   0.0240 & 19.783 &  0.053 \\
   0.0261 & 19.681 &  0.052 \\
   0.0296 & 19.938 &  0.057 \\
   0.0344 & 20.130 &  0.059 \\
   0.0393 & 20.291 &  0.063 \\
   0.0459 & 20.516 &  0.076 \\
   0.0542 & 20.833 &  0.094 \\
   0.0625 & 20.842 &  0.093 \\
   0.0749 & 21.127 &  0.130 \\
   0.0832 & 21.523 &  0.212 \\
\enddata
\label{060210_lightcurve}
\tablecomments{Table~\ref{060210_lightcurve} is available in its
entirety in the electronic version of the Journal. A portion is shown
here for guidance regarding its form and content.}
\end{deluxetable}

\end{document}